\documentstyle[12pt]{article}
\newcommand{\plb}{{\em Phys. Lett. B} }
\newcommand{\npb}{{\em Nucl. Phys. B}  }
\newcommand{\prd}{{\em Phys. Rev. D} }
\def\hybrid{\topmargin -20pt	\oddsidemargin 0pt
	\headheight 0pt	\headsep 0pt
	\textwidth 6.25in	
	\textheight 9.5in	
	\marginparwidth .875in
	\parskip 5pt plus 1pt	\jot = 1.5ex}

\hybrid
\newskip\humongous \humongous=0pt plus 1000pt minus 1000pt
\def\caja{\mathsurround=0pt}
\def\eqalign#1{\,\vcenter{\openup1\jot \caja
	\ialign{\strut \hfil$\displaystyle{##}$&$
	\displaystyle{{}##}$\hfil\crcr#1\crcr}}\,}
\newif\ifdtup

\relax

\begin{document}
\renewcommand{\theequation}{\thesection.\arabic{equation}}
\newcommand{\beq}{\begin{equation}}
\newcommand{\eeq}[1]{\label{#1}\end{equation}}
\newcommand{\ber}{\begin{eqnarray}}
\newcommand{\eer}[1]{\label{#1}\end{eqnarray}}
\begin{titlepage}
\begin{center}

\hfill HUB-IEP-94/18\\
\hfill hep-th/9409095\\

\vskip .9in

{\large \bf

Duality Symmetries and Supersymmetry Breaking in String
Compactifications\footnote{Invited talk
to the 27th ICHEP, Glasgow, July 1994, presented by D. L\"ust}
}
\vskip .9in

\vskip .15in

{\bf Gabriel Lopes Cardoso and Dieter L\"ust}\\
\vskip
 .1in

{\em Humboldt-Universit\"at zu Berlin\\
Institut f\"ur Physik\\
D-10099 Berlin} \footnote{e-mail addresses:
GCARDOSO@QFT2.PHYSIK.HU-BERLIN.DE, LUEST@QFT1.PHYSIK.HU-BERLIN.DE}\\

\vskip .3in
and
\vskip .3in

{\bf Thomas Mohaupt}
\vskip .1in

{\em  DESY-IfH Zeuthen\\
Platanenallee 6\\
D-15738 Zeuthen}\footnote{e-mail address: MOHAUPT@HADES.IFH.DE
}

\end{center}

\vskip .4in

\begin{center} {\bf ABSTRACT } \end{center}
\begin{quotation}\noindent
We discuss the spontaneous supersymetry breaking
within the low-energy effective supergravity action of four-dimensional
superstrings. In particular, we emphasize the non-universality
of the soft supersymmetry breaking parameters, the $\mu$-problem and
the duality symmetries.

\end{quotation}
\vskip 3.0cm
September 1994\\
\end{titlepage}
\vfill
\eject
\def\baselinestretch{1.2}
\baselineskip 16 pt
\setcounter{equation}{0}

\section{Introduction}

Based on theoretical motivations, in particular the socalled
hierarchy problem, and stimulated by some indirect experimental hints,
like coupling constant unification and the top quark mass, the
minimal supersymmetric standard model (MSSM) was extensively
discussed during the last years. Unfortunately, the necessary violation
of supersymmetry has to be put in by hand into the MSSM and is described
by the socalled soft supersymmetry breaking parameters (SSBP)
like the gaugino masses
etc. For reasons of simplicity these SSBP
were assumed in most of the phenomenological discussions to be universal
for all different gauginos and also for the various matter fields.
For some SSBP, a possible deviation from
universality is severely constrained by phenomenological requirements
like the absence a flavor changing neutral currents \cite{fcnc}.

On the other hand,
superstring theories are a very promising candidate for a consistent
quantization of gravity. For this purpose, the typical string scale has to
be identified with the Planck mass $M_P$ of order $10^{19}$ GeV.
Therefore one strongly hopes that superstrings may solve some puzzles
concerning quantum physics at $M_P$.
Now for the actual relevance of
superstring theories it is of most vital importance to make direct
contact  to the standard model
(SM) or perhaps better to the MSSM.
This programm attracted a lot of attention during the last 10 years,
and the results of this research are,
at least conceptually, quite successful.
Indeed, the low energy effective
lagrangian of a large class of four-dimensional heterotic string theories
is just given by the standard $N=1$ supergravity action
with gauge group potentially
containing the gauge group of the SM and with matter coming very near to
the three chiral families of the SM.
Deriving the effective string action,
it is very important to realize that the low energy
spectrum and the low energy effective interactions among the almost massless
fields are  to some extent controlled by the stringy symmetries which are
still reminiscent after integrating out the infinite number of massive modes.
A particular nice example of this kind are the well established duality
symmetries (for a review see
\cite{gpr}) which proved to provide useful information about the
effective string action on general grounds.

A very attractive feature of $N=1$ supergravity in general is the fact that
upon
spontaneous supersymmetry breaking in some hidden sector of the theory
the SSBP in the observable sector
automatically emerge due to gravitational couplings among observable and hidden
fields. Thus in string theory the SSBP are, at least in principle, calculable
from first principles. However at the moment,
the actual mechanism of supersymmetry breaking is far from being completely
understood.
However recently it was demonstrated \cite{il1,kl1,bim,fkz,kzp} that,
parametrizing the SSBP without specifying the actual supersymmetry
breaking mechanism,  some interesting generic features
of supersymmetry breaking in superstring theories can be derived.
In particular it turned out that the SSBP are generically non-universal
\cite{il1}.

This contribution will be organized as follows: first we will set up
the general formalism of supersymmetry breaking in $N=1$ supergravity
with special emphasis on
the structure of the SSBP in four-dimensional strings. As  more specific
examples we will then present some results for Abelian orbifolds.

\section{$N=1$ effective supergravity action for four-dimen\-sio\-nal heterotic
strings}

Let us first specify the string modes with masses small compared to
$M_P$ which we assume to appear in the effective action.
First there is the $N=1$ supergravity multiplet containing
the graviton field and the spin $3\over 2$ gravitino. Next,
the gauge degrees of freedom are described by $N=1$ vector  multiplets
$V_a$ with spin 1 gauge bosons and Spin $1\over 2$ gauginos $\lambda_a$.
The gauge index $a$ is assumed to range over the SM gauge group
$SU(3)\times SU(2)\times U(1)$ and an unspecified hidden gauge group
$G_{\rm hid}$.
Finally we consider chiral matter multiplets $\Phi^I$ with complex scalars
and spin $1\over 2$ Weyl fermions.
These chiral fields, i.e. the index $I$, separate into socalled
matter fields $Q^\alpha$ which contain the matter of the MSSM,
$Q_{\rm SM}^\alpha=(q,l,H_1,H_2)$, and matter which only transforms
non-trivilly
under $G_{\rm hid}$, $Q_{\rm hid}^\alpha$. The second type of chiral fields
$\Phi^I$
correspond to the socalled moduli fields $M^i$ whose vacuum expextation
values (vev's)
are undetermined
in perturbation theory since the $M^i$ correspond to the free parameters
of the four-dimensional string models.
The moduli are assumed to be SM singlets (however note that $H_1$ and $H_2$
could be in principle moduli).
The duality group $\Gamma$ acts on the moduli
$M^i$ as discrete reparametrizations, $M^i\rightarrow\tilde M^i(M^i)$,
which leave the underlying four-dimensional string theory invariant.
Therefore, the effective action of the massless field must be
$\Gamma$ invariant which provides a link between $L_{\rm eff}$ and the
theory of $\Gamma$-modular functions \cite{flst}. Moreover
strong restrictions on the massless spectrum arise \cite{il1}
due to the required
absence of potential duality anomalies.

The effective $N=1$ supergravity action, up to two space-time derivatives,
is specified by three different functions of the chiral fields $\Phi^I$
\cite{cfgp}.
First, the K\"ahler potential $K$ is a  gauge-invariant real analytic
function of the chiral superfields. To compute later on the SSBP it
is enough to expand $K$ up to quadratic order in the matter fields:
\begin{equation}
K=K_0(M,\bar M)+K_{\alpha\bar\beta}(M,\bar M)Q^\alpha \bar Q^{\bar\beta }
+({1\over 2}H_{\alpha\beta}(M,\bar M)Q^\alpha Q^\beta+{\rm h.c.})
\end{equation}
Note that for SM matter fields the last term in eq.(1) can be non-vanshing
only for a mixing term of the two Higgs fields: $({1\over 2}H_{12}(M,\bar M)
H_1H_2+{\rm h.c.})$. $K_0$ is just the K\"ahler potential of the
K\"ahlerian moduli space ${\cal K}_0$. $\Gamma$-duality transformations act as
K\"ahler transformations on $K$, $K\rightarrow K+g(M)+\bar g(\bar M)$
($g(M)$ is a holomorphic function of the moduli), and induce a `rotation'
on the matter fields, $Q^\alpha\rightarrow h_{\alpha\beta}(M)Q^\beta$.

Next we consider the moduli dependent effective gauge couplings $g_a(M,\bar
M)$:
\begin{equation}
\eqalign{g_a^{-2}(M,\bar M)&={\rm Re}~f_a(M)
 -{1\over 16\pi^2}\biggl((C(G_a)-\sum_\alpha T_a(\alpha))K_0(M,\bar M)
\cr &+2\sum_\alpha T_a(\alpha)\log\det K_{\alpha\bar\beta}(M,\bar
M)\biggr).\cr}
\end{equation}
$C(G_a)$ is the quadratic Casimir of the gauge group $G_a$ and
$T_a(\alpha)$ the index of the massless matter representations.
The holomorphic gauge kinetic function $f_a(M)$ includes the
tree level moduli dependence as well as possible one-loop quantum corrections
from massive modes; however beyond one loop the are no perturbative
corrections to $f_a(M)$ \cite{nkl}.
The non-holomorphic terms in eq.(2) originate
from one-loop corrections involving massless fields. Specifically,
these terms describe the presence of K\"ahler as well as $\sigma$-model
anomalies \cite{anom,dfkz}.
$g_a(M,\bar M)$ has to be a duality invariant function. Therefore
the duality non-invariance of the non-holomorphic anomaly terms
has to be cancelled by a non-trivial transformation behaviour
of
$f_a(M)$:
\begin{equation}
f_a(M)\rightarrow f_a(M)+{1\over 8\pi^2}((C(G_a)
-\sum T_a(\alpha))
g(M)-{1\over 4\pi^2}\sum T_a(\alpha)\log\det h_{\alpha\beta}(M).
\end{equation}

Third the superpotential will be conveniently split into a SUSY-preserving
tree-level part and into a SUSY-breaking piece which does not
depend on the matter fields:
\begin{equation}
W=W_{\rm tree}(Q,M)+W_{\rm SUSY-breaking}(M).
\end{equation}
Duality invariance of the effective action demands that $W$ transforms as
$W\rightarrow e^{-g(M)}W$.
The structure of $W_{\rm tree}$ is such that it generates
the moduli-dependent  Yukawa couplings for the matter fields as well
as possible moduli-dependent mass terms for some hidden matter fields;
the observed matter fields are assumed to stay massless for all
values of the moduli fields:
\begin{equation}
W_{\rm tree}={1\over 3}h_{\alpha\beta\gamma}(M^i)Q^\alpha Q^\beta Q^\gamma+
{1\over 3}h_{i\alpha\beta}(M^i) Q^\alpha_{\rm hid}Q^\beta_{\rm hid}.
\end{equation}
Thus it may happen that at some points in the moduli space, $h_{i\alpha
\beta}(M^i)=0$,
there are additional massless hidden matter fields. Very often they
go together with additional massless gauge bosons at these points.

Essentially, there are two very promising mechanisms of
supersymmetry breaking in the last years' literature. First at tree level by
the socalled Scherk-Schwarz mechanism \cite{sw}. This can be described
in the effective field theory by a  tree-level superpotential.
Second supersymmetry can be broken due to
non-perturbative effects.
Unfortunately it is not possible at the moment to calulate these
non-perturbative effects directly in string theory.
However, let us assume that
non-perturbative field theory effects give a dominant
contribution to the spontaneous breaking of supersymmetry. In particular,
one can show that non-perturbative gaugino condensation in the hidden gauge
sector  potentially breaks supersymmetry \cite{gc}. Integrating out the
dynamical degrees of freedom corresponding to the gaugino
bound states, the
duality invariant \cite{filq,fmtv} gaugino condensation can be described
by an effective non-perturbative
superpotential, which depends holomorphically on the moduli fields:
\begin{equation}
W_{\rm SUSY-breaking}(M)=e^{{24\pi^2\over b_a}f_a(M)},
\end{equation}
($b_a$ is the $N=1$ $\beta$-function coefficient).
It is remarkable that this expression is in a sense exact since
$f_a(M)$ is only renormalized up to one loop.
It is this exactness of $W_{\rm SUSY-breaking}$ which
provides very strong confidence in the
applicability of the used method.

Now let us discuss the form of the SSBP in the effective action
which arise after
the spontaneous breaking of local supersymmetry. This discussion will not
refer to the actual (perturbative or non-perturbative) breaking mechanism;
nevertheless some interesting information about these
couplings can be obtained at the end.
The scalar potential in the low-energy
supergravity action has the form \cite{cfgp}
\begin{equation}
V=|W_{\rm SUSY-breaking}(M)|^2e^{K_0}(G^iG_i-3).
\end{equation}
($e^G=|W|^2e^K$, $G_I={\partial G\over\partial\phi_I}$.) Deriving
this formula we have assumed that, upon minimization of $V$,
$<G_\alpha >=0$ and $<Q_\alpha >=0$ in the matter sector. This assumption,
which is satisfied in most realistic scenarios, means that the spontaneous
supersymmetry breaking takes places in the moduli sector, i.e. $<G_i>\neq 0$
for at least one of the moduli fields. Then the gravitino mass becomes
\begin{equation}
m_{3/2}=e^{K_0(M,\bar M)/2}|W_{\rm SUSY-breaking}(M)|.
\end{equation}
$m_{3/2}$ should be of order TeV; thus the smallness of this scale compared
to $M_P$ must come either from the K\"ahler potential and/or from
the superpotential.
Now we obtain the following SSBP: first the gaugino masses
take the form
\begin{equation}
m_a(M\,\bar M)={1\over 2}m_{3/2}G^i(M,\bar M)\partial_i\log g_a^{-2}(M,\bar M).
\end{equation}
The scalar masses (squarks and sleptons) become \cite{kl1}
\begin{equation}
m^2_{\alpha\bar\beta}=m^2_{3/2}\lbrack K_{\alpha\bar\beta}(M,\bar M)-
G^i(M,\bar M)G^{\bar j}(M,\bar M)R_{i\bar j\alpha\bar \beta }\rbrack .
\end{equation}
($R_{i\bar j\alpha\bar\beta}=\partial_i\bar\partial_{\bar j}K_{\alpha\bar\beta}
-\Gamma_{i\alpha}^\gamma K_{\gamma\bar\delta}\bar\Gamma_{\bar j\bar\beta}^{
\bar\delta}$, $\Gamma_{i\alpha}^\gamma=K^{\gamma\bar\delta}\partial_iK_{\alpha
\bar\gamma}$. These parameters are generically of the order of $m_{3/2}$.
Their exact values depend on the details of $K$, $W$ and the (dynamically
fixed) vev's of the moduli fields. It is quite evident
that in general these SSBP are non-universal \cite{il1}.
The non-universality arises due to the
non-universal moduli dependence of the
gauge couplings and the matter kinetic energies. Similar expression can be also
obtained for the trilinear couplings \cite{cfilq,kl1,bim}

Finally let us investigate the possible apperance of a mass mixing term
for the two standard model Higgs fields $H_1$, $H_2$ which is necessary
for the correct radiative breaking of the electro-weak gauge symmetry.
Clearly, a tree-level mixing
due to a quadratic term in the superpotential,
$W_{\rm tree}=\mu H_1H_2$, would be a desaster, since it will be
most likely of the order of $M_P$.
(This is often called the $\mu$-problem.) However, if there exist
\cite{gm} a possible,
holomorphic
mixing term $H_{12}$ among $H_1$ and $H_2$ in the tree-level K\"ahler potential
(see eq.(1)), then an effective $\mu$-term will be generated after
the spontaneous breaking of supersymmetry:
\begin{equation}
W_{\rm eff}=\hat\mu H_1H_2,\qquad
\hat\mu=m_{3/2}
\lbrack H_{12}(M,\bar M)-G^{\bar i}\partial_{\bar i}H_{12}(M,\bar M)\rbrack .
\end{equation}

\section{Abelian orbifolds}

In this chapter we want to apply our previous formulas to the
case of  Abelian orbifold compactifications \cite{dhvw}.
Every orbifold of this type
has three complex `planes', and each orbifold twist $\vec\theta=\theta_i$
(i=1,2,3) acts either simultaneously on two or
all three planes. Generically, for all four-dimensional strings there exist
as moduli fields the dilaton ($D$) -- axion ($a$) chiral multiplet  $S=e^D+ia$.
Then the tree-level K\"ahler potential
for the $S$-field has the form $K_0=-\log(S+\bar S)$.

Next we consider the internal moduli of the orbifold compactification.
We will concentrate on the untwisted moduli fields. For each Abelian
orbifold there exist at least three K\"ahler class moduli $T_i$ each
associated to one of the three complex planes.
We will call the $T_i$ (2,2) moduli, since they do not
destroy a possible (2,2) superconformal structure of the underlying
string theory, i.e. their vev's do not
break the (2,2) gauge group $E_6\times E_8$.
Next we consider socalled (0,2) untwisted moduli which are generically present
in any orbifold compactification. A non-vanishing vev
for these kind of fields destroys the (2,2) world sheet supersymmetry and
breaks $E_6\times E_8$ to some non-Abelian subgroup. In addition they
will generically give mass to some matter fields by a superpotential coupling.
Specifically these types of moduli correspond to continuous
Wilson line background fields \cite{inq} which are again associated to each of
the three complex planes.  For the case that $\theta_i\neq\pm 1$, there is
generically at least
one complex Wilson line field $A_i$ (for example a ${\underline
{27}}$ of $E_6$).
The combined $T_i$, $A_i$ K\"ahler potential reads \cite{ms,llm}
\begin{equation}
K_0=-\log(T_i+\bar T_i-A_i\bar A_i)
\end{equation}
 and leads to the K\"ahler metric of the
space ${\cal K}_0=SU(1,2)/SU(2)\times U(1)$.
If $\theta_i=\pm 1$
there will be additional moduli fields namely, first, the
(2,2) modulus $U_i$ which
corresponds to the possible deformations of the complex structure. In addition
there will be
again some
(0,2) moduli, namely generically at least
two
complex Wilson line moduli $B$ and $C$
\cite{llm}.  ($B$ and $C$ being, for example,
${\underline{27}}$ respectively ${\underline{\bar{27}}}$ of $E_6$).
Then the K\"ahler potential for these fields can be determined as follows
\cite{llm}:
\begin{equation}
K_0=-\log\lbrack (T_i+\bar T_i)(U_i+\bar U_i)-
{1\over 2}(B_i+\bar C_i)(\bar B_i+C_i)\rbrack .
\end{equation}
The corresponding K\"ahler moduli space is given by ${\cal K}_0=SO(2,4)/
SO(2)\times SO(4)$. A few remarks are at hand. First note that in the
absence of Wilson lines ($B=C=0$) the K\"ahler potential splits into
the sum $K_0=K(T,\bar T)+K(U,\bar U)$, which is the well-known
K\"ahler potential for the factorizable coset $SO(2,2)/SO(2)\times SO(2)=
SU(1,1)/U(1)_T\otimes SU(1,1)/U(1)_U$. On the other hand, turning on
Wilson lines, the moduli space does not
factorize anymore into two submanifolds. Thus it is natural to expect that also
in a more general situation the moduli space is not anymore factorizable
into a space of the K\"ahler class moduli times a space of the complex
structure moduli
(as it is true for (2,2) compactifications) as soon as (0,2) moduli are turned
on. Also note that the complex Wilson lines give rise to holomorphic
$BC$ and antiholomorphic $\bar B\bar C$ terms in the K\"ahler potential.
This is in principle just what is needed for the solution of the
$\mu$-problem; upon identification of $H_1$ with $B$ and $H_2$ with $C$ the
mass
mixing term becomes \cite{agnt,llm}
\begin{equation}
H_{12}={1\over (T+\bar T)(U+\bar U)}.
\end{equation}
(This is also true
in general if $B$ and $C$ are not moduli but matter fields with
tree-level zero vev's \cite{kl1,agnt}.) Thus we learn that holomorphic
mixing terms in the K\"ahler potential can occur only if $\theta_i=\pm 1$,
i.e. if there exists a complex structure modulus $U_i$. Consequently, the
Higgs fields should be associated to this particular complex plane.

Now let us briefly discuss the duality symmetries. We consider the most
interesting case with four complex moduli $T$, $U$, $B$ and $C$. (For
more discussion see \cite{llm}.) In addition we assume that the complex plane
corresponds to a two-dimensional subtorus. The duality, i.e. modular group
in question is then given by the discrete group $O(2,4,Z)$.
The modular group $O(2,4,Z)$ contains an $SO(2,2,Z)=
PSL(2,Z)_T\times PSL(2,Z)_U$ subgroup.  $PSL(2,Z)_T$ acts in the standard
way on the $T$ modulus
\begin{equation}
T\rightarrow{aT-ib\over icT+d}
\end{equation}
($a,b,c,d\in Z$,
$ad-bc=1$). However $U$ transforms also non-trivially under this transformation
as
\begin{equation}
U\rightarrow U-{ic\over 2}{BC\over icT+d}.
\end{equation}
Thus, in the presence of $B$ and
$C$,
$T$ and $U$ get mixed under duality transformations  \cite{llm,agnt}
which reflects
the non-factorizable structure of the moduli space.

For the discussion of supersymmery breaking one also needs to
include one-loop corrections to the moduli K\"ahler potential. These
arise
due to a one-loop mixing of the $S$-field with the internal moduli.
This  is the socalled Green-Schwarz mixing with mixing coefficient
$\delta_{GS}^i$.
Specifically one can show that the loop corrected K\"ahler potential
has the following structure
\cite{dfkz,loop}:
\begin{equation}
\eqalign{&
K_0^{\rm 1-loop}=-\log Y+K_0^{\rm tree}(
T,U, A, B, C), \cr &Y=S+\bar S+{1\over 8\pi^2}\sum_{i=1}^3
\delta_{GS}^iK_{0~i~{\rm tree}}(T_i,U_i,A_i,B_i,C_i).\cr}
\end{equation}
Furthermore for the computation of the SSBP we need the tree-level
K\"ahler potential of the matter fields. It can be shown to have the following
form
\cite{dkl1,il1}:
\begin{equation}
K_{\alpha\bar\beta}=\delta_{\alpha\bar\beta}\prod_{i=1}^3(T_i+\bar T_i)^
{n_\alpha^i}.
\end{equation}
(For simplicity we have included only the generic $T_i$
moduli.) The integers $n_\alpha^i$ are called modular weights of the
matter fields, since the $Q_\alpha$ transform under $PSL(2,Z)$ as
\begin{equation}
Q^\alpha
\rightarrow Q^\alpha\prod_{i=1}^3(ic_iT_i+d_i)^{n_\alpha^i}.
\end{equation}
(The Wilson
line moduli $A,B,C$ have modular weight -1.)

As a final ingredient we have to specify the form of the gauge kinetic
function in orbifold compactifications. Including one $A$-type modulus,
the $f$-function in lowest order in $A$ is given as
\begin{equation}
f(S,T_i,A)_a=S-{1\over 8\pi^2}(b_1-b_0)\log \lbrack h(T_i)A\rbrack
 -{1\over 8\pi^2}\sum_{i=1}^3({b'}^i_a-\delta_{GS}^i)\log\eta(T_i)^2
{}.
\end{equation}
Here $\eta(T_i)$ is the well-known Dedekind function
and reflects the one-loop threshold contributions of
momentum and winding states \cite{th}. The $A$ contribution corresponds to the
mass thresholds \cite{mth1,mth2}
of those fields $Q^\alpha $
which get mass by a superpotential
coupling to $A$: $W\sim h(T_i)AQ^\alpha Q^\beta$ . If one assumes that
all matter fields, that are charged under $G_a$, get a $A$-dependent masses
one obtains $b_0=-3C(G_a)$,
$b_1=-3C(G_a)+\sum_{\alpha }T_a(\alpha )$. Then ${b'}_a^i=-C(G_a)+
\sum_\alpha T_a(\alpha )(1+2n_\alpha^i)$. It is not difficult
to verify the correct duality transformation behaviour of $f$.

Now let us apply these formulas to discuss some specific aspects of
supersymmetry breaking in orbifold compactifications.
Let us  focus  on the non-perturbative gaugino condensation in
the hidden gauge sector $a$. The
non-perturbative superpotential then reads
\begin{equation}
W_{\rm SUSY-breaking}={e^{{24\pi^2\over b_0}S}\lbrack h(T_i)
A\rbrack^{3(b_0-b_1)/b_0}\over
\prod_{i=1}^3\lbrack\eta(T_i)\rbrack^{6({b'}^i_a-\delta_{GS}^i)/b_0}}
{}.
\end{equation}
This leads to the following
expression \cite{mth2} for the scalar potential $V$ using
the one-loop corrected K\"ahler potential but neglecting
for simplicity a possible $A$ contribution, i.e. $b_0=b_1=3{b'}_a^i$
(the inclusion of $A$ can be found in \cite{lt,mth2}):
\begin{equation}
V=m_{3/2}^2\biggl\lbrace |1-{24\pi^2\over b_0}Y|^2+\sum_{i=1}^3{Y\over
8\pi^2Y-\delta_{GS}^i}
(1-3{\delta_{GS}^i\over b_0})(T_i+\bar T_i)^2|\hat G_2(T_i)|^2-3\biggr\rbrace
{}.
\end{equation}
The minimization of this scalar potential leads to the following results.
First note that in case of complete Green-Schwarz cancellation,
i.e. $b_0=3\delta_{GS}^i$, there is no $T_i$ dependence in the potential
(as well as in $m_{3/2}$) and $T_i$ still remains as a undetermined parameter.
On the other hand, for
$3\delta_{GS}^i\neq b_0$, the modulus $T_i$ gets dynamically fixed.
A specific analysis was performed in \cite{filq,cfilq}
for the case $\delta_{GS}^i=0$
with the result that at the minimum $T_i\sim 1.2$  supersymmetry
gets spontaneously broken in the $T_i$ sector since at that point $G_{T_i}
\neq 0$. However there is an important caveat witin this analysis since
it used the assumption that at the minimum $G_S=0$. In fact,  the above
potential, triggered by the gaugino condensate, has no stable
minimum with respect to $S$. Therefore the dilaton dynamics has to be modified
in order to justify this assumption. One way could be that there
are gaugino condensates in more that one hidden gauge sector \cite{twog}. Then
$G_S=0$ is rather generic, however several
$\beta$-function coefficients have to be
tuned in a careful way in
order to get $m_{3/2}\sim$O(1TeV). A different, very interesting possibility
is that the non-perturbative dilaton dynamics is governed by the
socalled $S$-duality \cite{sdual1,sdual2}.
This means that the true non-perturbative string
partition function is actually  $PSL(2,Z)$ invariant resp. covariant
with respect to the $S$-field due to non-perturbative monopol-like
configuration in target space. The simplest possibility within this
context is that the partition function looks like
\cite{sdual1}
\begin{equation}
Z\sim{1\over (S+\bar S)
|\eta(S)|^4}
{}.
\end{equation}
In the effective field theory this could mean that the effective
superpotential contains a term $\eta(S)^{-2}$ instead of the `standard'
$e^S$ dependence. Such types of superpotentials possibly lead to $G_S=0$.
Finally one has to remark in this context that the cosmological constant
tends to be non-vanishing within the non-perturbative scenario, which
is very disturbing but probably reflects our ignorance about
the exact supersymmetry breaking dynamics dynamics. (For a recent discussion
about the cosmological constant see \cite{fkz}; in \cite{ckn} it has
been argued that a negative cosmological constant after gaugino condensation
might be a desirable feature, for the fully renormalized cosmological constant
to vanish.)

Now, we  could proceed to calculate the SSBP resulting
from this type of superpotentials.
For example the squark and slepton masses are
obtained as a function of the modular weights $n_\alpha$ \cite{il1}.
At this stage
it is very
convenient to parametrize the unknown supersymmetry dynamics
by some angle $\tan\theta\sim{G_S\over G_T}$ \cite{bim},
i.e. the relative strength of the supersymmetry breaking in the $S$
and $T$ sectors. Then
the exact form of the (perturbative or non-perturbative)
superpotential is parametrized by $\theta$ and $m_{3/2}$, and
the form of the SSBP depends only on  known perturbative
quantities like $K$.  Specifically the scalar masses
have the form (assuming vanishing cosmological constant, the index
$i$ is suppressed now) \cite{bim}:
\begin{equation}
m_\alpha^2=m_{3/2}^2\lbrack 1+n_\alpha
(1-{\delta_{GS}\over 24\pi^2 Y})^{-1}\cos^2\theta\rbrack.
\end{equation}
For arbitrary values of $\theta$ these SSBP are non-universal.
However for $\theta=\pi/2$, i.e. the dilaton dominated supersymmetry
breaking, the SSBP are in fact universal \cite{blm}.
Finally, for the gaugino masses similar expressions
can be derived. Concluding, it would be very interesting to test some of these
features in future colliders.

\end{document}